# Searching for Single Production of Charged Heavy Leptons via Anomalous Interactions at CLIC

A. T. Tasci[*] and A. Senol[†]

*Kastamonu University, Department of Physics,*

*37100, Kuzeykent, Kastamonu, Turkey*

## Abstract

We consider the possible discovery potential for single production of charged heavy leptons via anomalous interactions at the envisaged Compact Linear Collider (CLIC) by taking into account initial state radiation (ISR) and beamstrahlung effects. We calculate the production cross sections and decay widths of charged heavy leptons in the context of anomalous interactions at center of mass energies $\sqrt{s}$ = 1 and 3 TeV. The signal and corresponding backgrounds are studied in detail for the mass range 300-1900 GeV.

[*]atasci@kastamonu.edu.tr

[†]asenol@kastamonu.edu.tr

# I. INTRODUCTION

At the present time, the experimental evidence for new physics beyond the Standard Model (SM) is insufficient. The observation of new heavy particles would change this situation completely. Thus, any signal for the production of heavy leptons will play a milestone role in the discovery of new physics. Although, we expect many exciting discoveries with start-up of the Large Hadron Collider (LHC), which will search for TeV scale energies and masses, heavy lepton production at the LHC would be difficult to detect, due to large backgrounds and small production rate. A linear collider with clean experimental environment, can also provide complementary information for new physics with performing precision measurements that would complete the LHC results. A high enregy linear collider can provide the best limits on their masses and mixings. One of the proposed linear collider with energies on the TeV scale and extremely high luminosity is the Compact Linear Collider (CLIC) ($\sqrt{s}$ = 1 and 3 TeV, $L^{int}$ = $10^5$ pb$^{-1}$ per year) [1].

Although there are lots of studies on the new heavy leptons in the literature, namely, at future $e^-e^+$ [2–6], hadron [7–12] and $ep$ colliders [13, 14], further study via anomalous interactions is needed, since an anomalous coupling, which has the formalism in [15], that generalize the SM as effective operators of dimension greater than four, can have a considerable effect on the heavy lepton production especially at future colliders with their high energies and luminosities. In Refs. [16] and [17] it has been studied the single production of the fourth SM generation charged leptons via FCNC interaction with dimension-5 anomalous couplings at $ep$ and $\gamma e$ colliders, respectively. Any observation of these anomalous interactions would be the signal of new physics beyond the SM. Therefore, in the present work, we study possible production and decay processes of the charged new heavy leptons ($L$) via dimension-5 anomalous magnetic moment type interactions in $e^-e^+$ collision including initial state radiation (ISR) and beamstrahlung effects. We also calculate the discovery bounds of masses for heavy leptons with anomalous couplings.

The experimental upper bounds for the heavy lepton masses were found to be 81.5 GeV by OPAL [18], 63.5 GeV by ALEPH [19], 225 GeV by H1 [20] and 101.9 GeV by L3 [21] Collaborations. Considering these experimental upper bounds, we scan the mass range of heavy leptons between 300-1900 GeV at the envisaged TeV energies of the CLIC.

# II. SINGLE PRODUCTION AND DECAYS OF HEAVY LEPTON

The model independent effective Lagrangian having magnetic moment type operators that describes the anomalous interactions of heavy leptons, by ordinary ones can be written from [22] with minor modifications as:

$$\mathcal{L}_{eff} = ie\frac{\kappa_\gamma^{l_j}}{\Lambda}\bar{L}\sigma_{\mu\nu}q^\nu l_j A^\mu + \frac{ig}{2cos\theta_W}\frac{\kappa_Z^{l_j}}{\Lambda}\bar{L}\sigma_{\mu\nu}q^\nu l_j Z^\mu + \frac{ig}{\sqrt{2}}\frac{\kappa_W^{l_j}}{\Lambda}\bar{L}\sigma_{\mu\nu}q^\nu P_L \nu_{l_j} W^\mu + h.c., \qquad (1)$$

where $j$=1, 2, 3 refers to the SM generation numbers; $\kappa_\gamma^l$, $\kappa_Z^l$ and $\kappa_W^l$ are the anomalous magnetic dipole moment factors (in this study $\kappa_{\gamma,Z,W}^{l_j} = 0$ when $j \neq 1$), $q$ is the momentum of the exchanged gauge boson, $\theta_W$ is the Weinberg angle, $e$ and $g$ denote the gauge couplings relative to U(1) and SU(2) symmetries respectively; $A^\mu$, $Z^\mu$ and $W^\mu$ are the vector fields of the photon, $Z$-boson and $W$-boson, respectively, $P_L$ is the left-handed projection operator and $\Lambda$ is the cutoff scale for new physics. The single production of heavy lepton occurs via the $s$ and $t$-channel and $Z$-boson exchange $e^-e^+ \to Le^+$ Flavor Changing Neutral Current (FCNC) process in $e^-e^+$ collision. Corresponding Feynman diagrams for production and their subsequent anomalous FCNC decays are shown in Fig. 1.

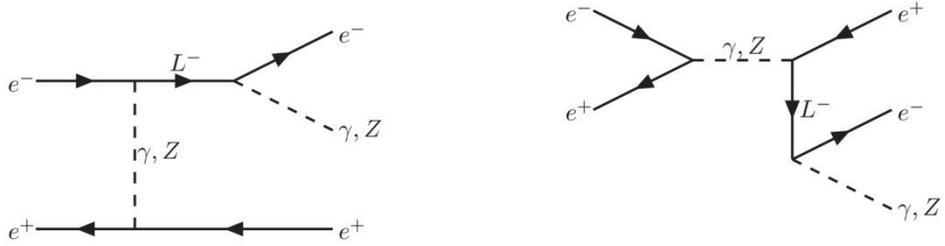

**FIG. 1:** Feynman diagrams relevant for single production of $L$ in $e^-e^+$ collisions.

The total differential cross section of this process is given by,

$$\frac{d\sigma}{dt} = \left(\frac{1}{128\pi s^2}\right)\Bigg\{\frac{16e^4}{st}\left(\frac{\kappa_\gamma^l}{\Lambda}\right)^2 \left(2(s^2+t^2)m_L^2 - (s+t)m_L^4 - 2(s^3+t^3)\right)$$

$$+ \frac{e^4(a_l^2+v_l^2)}{\sin^4\theta_W \cos^4\theta_W}\left(\frac{\kappa_Z^l}{\Lambda}\right)^2 \Bigg[\frac{\left((s+2t)m_L^2 - m_L^4 - 2t(s+t)\right)s}{(s-M_Z^2)^2 + M_Z^2\Gamma_Z^2}$$

$$+ \frac{\left((2s+t)m_L^2 - m_L^4 - 2s(s+t)\right)t}{(t-M_Z^2)^2 + M_Z^2\Gamma_Z^2}$$

$$+ \frac{2st(s+t-m_L^2)(st+M_Z^4 - (s+t-\Gamma_Z^2)M_Z^2)}{((s-M_Z^2)^2 + M_Z^2\Gamma_Z^2)((t-M_Z^2)^2 + M_Z^2\Gamma_Z^2)}\Bigg]$$

$$+ \frac{8e^4 v_l \Gamma_Z M_Z}{\sin^2\theta_W \cos^2\theta_W((s-M_Z^2)^2 + M_Z^2\Gamma_Z^2)((t-M_Z^2)^2 + M_Z^2\Gamma_Z^2)}\left(\frac{\kappa_\gamma^l}{\Lambda}\right)\left(\frac{\kappa_Z^l}{\Lambda}\right)[(s^2$$

$$+ t^2)m_L^4 - 2(s^3+t^3)m_L^2 + (s+t)^2(2s^2 - 3st + 2t^2)$$

$$+ ((2m_L^4 - 2(s+t)m_L^2 + (s+t)^2)(M_Z^2 + \Gamma_Z^2)$$

$$- 2\left((s+t)m_L^4 - 2(s^2+t^2)m_L^2 + 2(s^3+t^3)\right))M_Z^2]\Bigg\} \qquad (2)$$

Here, $v_l = -1/2 + 2\sin^2\theta_W$, $a_l = -1/2$, $m_L$ is the mass of heavy lepton, $s$ and $t$ are the Mandelstam variable, $\Gamma_Z$ and $M_Z$ are decay width and mass of the Z-boson, respectively.

For the calculation of $e^- e^+ \to e^- \gamma(Z) e^+$ cross section including anomalous FCNC contributions, we need the total decay widths ($\Gamma_Z$) of heavy lepton. Heavy leptons decay via charged and neutral currents through the processes $L \to \gamma l$, $L \to Zl$ and $L \to W \bar{\nu}_l$. Neglecting ordinary lepton masses the decay widths are obtained as:

$$\Gamma(L \to \gamma l) = \frac{e^2}{8\pi}\left(\frac{\kappa_\gamma^l}{\Lambda}\right)^2 m_L^3, \qquad (3)$$

$$\Gamma(L \to Zl) = \frac{e^2}{64\pi \sin^2\theta_W \cos^2\theta_W}\left(\frac{\kappa_Z^l}{\Lambda}\right)^2 \left(2m_L^3 - 3M_Z^2 m_L + \frac{M_Z^6}{m_L^3}\right), \qquad (4)$$

$$\Gamma(L \to W\bar{\nu}_l) = \frac{e^2}{16\pi \sin^2\theta_W}\left(\frac{\kappa_W^l}{\Lambda}\right)^2 \left(2m_L^3 - 3M_W^2 m_L + \frac{M_W^6}{m_L^3}\right). \qquad (5)$$

### III. NUMERICAL ANALYSIS

In order to search for potential discovery of heavy leptons at CLIC, the anomalous vertices given in Eq. (1) are implemented into the tree-level event generator CompHEP [23].

Since the higher scale physics responsible for dimension-5 operator will be resolved at high energies, we assume $(\kappa/\Lambda) = 0.5(0.3)$ TeV$^{-1}$ for $\sqrt{s} = 1(3)$ TeV and in our numerical calculations by taking anomalous magnetic dipole moment factors as $\kappa_\gamma^l = \kappa_Z^l = \kappa_W^l = \kappa$. In Table I we present the total decay widths ($\Gamma_L$) of heavy lepton for the mass range 300-1900 GeV. Heavy lepton branching fractions remain constant at 15% ($L \to \gamma l$), 21% ($L \to Zl$) and 64% ($L \to W\bar{\nu}_e$) for the given mass range.

**TABLE I:** Total decay widths of heavy leptons ($\Gamma_L$) depending on its mass values for $\kappa = 0.5(0.3)$ TeV$^{-1}$.

| $m_L$ (GeV) | $\Gamma_L$ (GeV) |
|---|---|
| 300 | 0.48 (0.17) |
| 400 | 1.19 (0.43) |
| 500 | 2.38 (0.86) |
| 600 | 4.16 (1.49) |
| 700 | 6.65 (2.39) |
| 800 | 9.97 (3.59) |
| 900 | 14.24 (5.12) |
| 1100 | 26.09 (9.39) |
| 1500 | 66.37 (23.89) |
| 1900 | 135.08 (48.63) |

Photon radiation from incoming electrons and positrons is called as the Initial State Radiation (ISR). We take into account this spectrum in our calculations by using the CompHEP program with beamstrahlung spectra. Beamstrahlung is a process of energy loss by the incoming electron due to its interaction with the positron (electron) bunch moving in the opposite direction. Beamstrahlung spectrum, which depends on the bunch geometry, bunch charge and the collision energy, is an attribute of the linear collider design. When calculating these effects we take into account the beam parameters of the CLIC as shown in Table II [1].

**TABLE II:** The beam parameters of the two energy options of the CLIC. N is the number of particles in the electron bunch, $\sigma_{x,y,z}$ are the average sizes of the electron bunches, $\Upsilon$ is the beamstrahlung parameter and $N_\gamma$ is the average number of photons per electron.

| Collider | CLIC | |
|---|---|---|
| Parameter | 1 TeV | 3 TeV |
| $N(10^{10})$ | 0.4 | 0.4 |
| $\sigma_x$(nm) | 115 | 43 |
| $\sigma_y$(nm) | 1.75 | 1 |
| $\sigma_z$(nm) | 30 | 30 |
| $\Upsilon$ | 1.014 | 8.068 |
| $N_\gamma$ | 1.04 | 1.74 |

In Figs. 2 and 3, we plot the production cross sections as function of $m_L$ for center of mass energies 1 TeV and 3 TeV, respectively, with and without ISR and beamstrahlung effects.

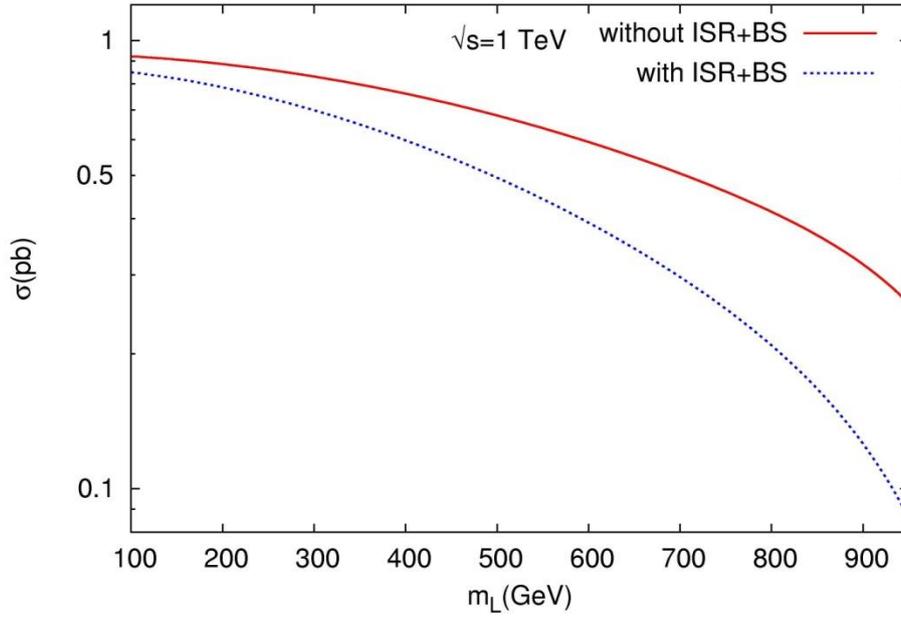

**FIG. 2:** The total cross sections for the process $e^- e^+ \to L^- e^+$, as function of $m_L$ for $\sqrt{s}$ = 1 TeV with and without ISR and beamstrahlung (BS) effects with $\kappa/\Lambda$ = 0.5 TeV$^{-1}$.

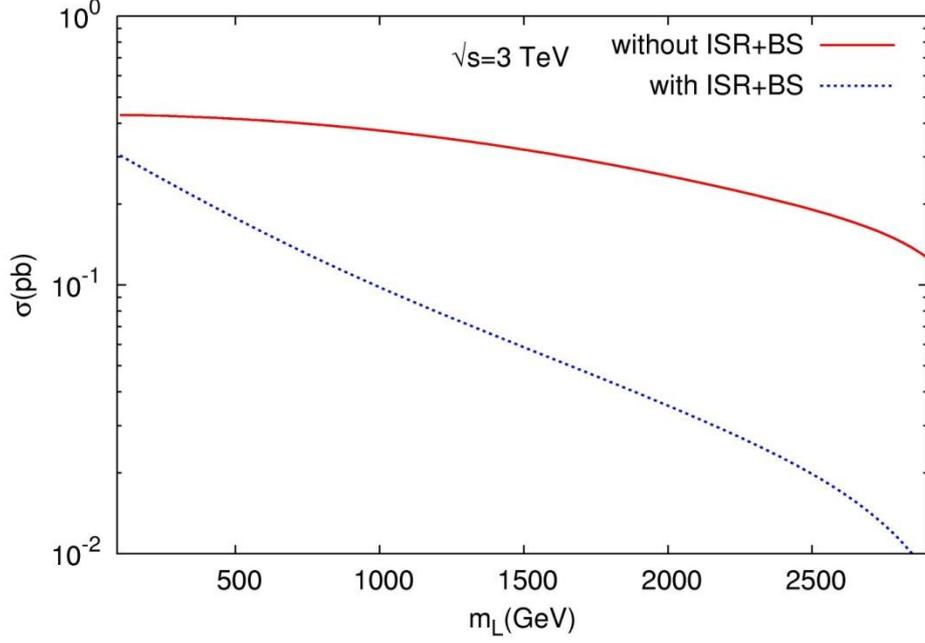

**FIG. 3:** The total cross sections for the process $e^-e^+ \to L^-e^+$, as function of $m_L$ for $\sqrt{s}$ = 3 TeV with and without ISR and beamstrahlung (BS) effects with $\kappa/\Lambda$ = 0.3 TeV$^{-1}$.

In this study, we assume that heavy leptons are produced via signal processes $e^-e^+ \to L^-e^+$. After their production, we consider, heavy leptons will decay into an electron and a photon or a Z boson only via FCNC interactions, that is, $e^-e^+ \to e^-\gamma e^+$ and $e^-e^+ \to e^-Ze^+$. The main SM background comes from the same final state processes as $e^-e^+ \to e^-\gamma e^+$ and $e^-e^+ \to e^-Ze^+$.

The signal is selected by cutting on the distribution of kinematical observables that can separate the signal from the SM backgrounds. In order to extract the signal and suppress the background, we apply the transverse momentum, $p_T^{e^\pm,\gamma} > 50$ GeV, and pseudo-rapidity, $|\eta^{e^\pm,\gamma}| < 2.5$, cuts. For realistic analysis, we take into account the finite energy resolution of electromagnetic calorimeter of the CLIC as $\delta E/E = 0.1/\sqrt{E} + 0.01$ [1]. The corresponding $\gamma(Z)e$ invariant mass resolution is given approximately by $\delta m_{\gamma(Z)e} = 0.1\sqrt{m_{\gamma(Z)e}} + 0.007 m_{\gamma(Z)e}$. In our calculations we use the mass bin width $\Delta m = max(\Gamma_L, \delta m_{\gamma(Z)e})$ to count signal and background events with the mass resolution $\delta m_{\gamma(Z)e}$. We display the differential invariant mass distributions of the SM background and the signal with different mass values of $m_L$ at the final states $\gamma e$ and $Ze$ systems, including the ISR and beamstrahlung effects for $\sqrt{s}$ = 1 TeV in Figs. 4 and 5. Figs. 6 and 7 shows the same distributions for $\sqrt{s}$ = 3 TeV.

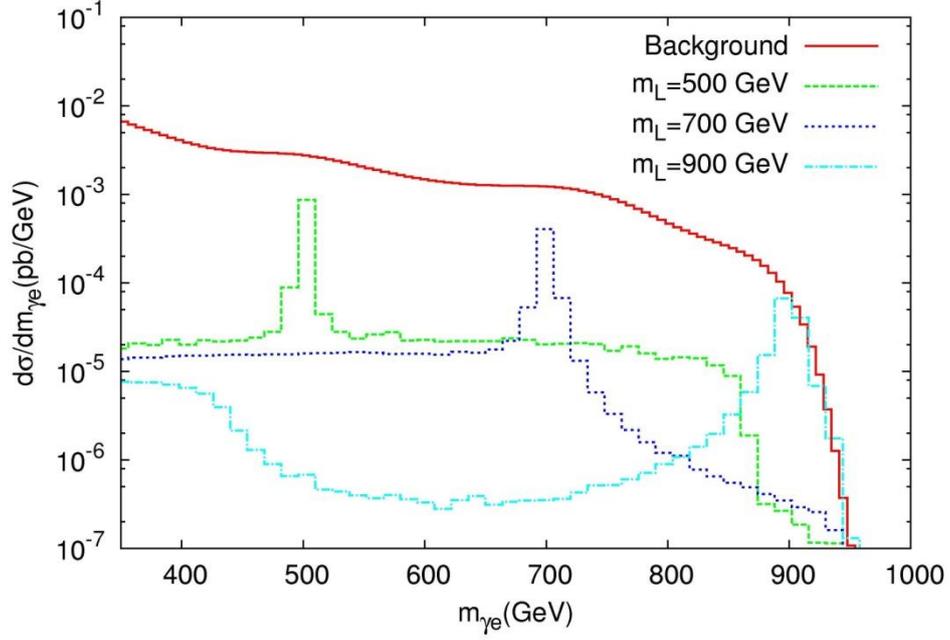

**FIG. 4:** The invariant mass distribution of $\gamma e$ in the final state at $\sqrt{s}$ = 1 TeV; signal with $m_L$ = 500 GeV, 700 GeV, 900 GeV and for the background ($e^-e^+ \to e^-\gamma e^+$) with $\kappa/\Lambda$ = 0.5 TeV$^{-1}$.

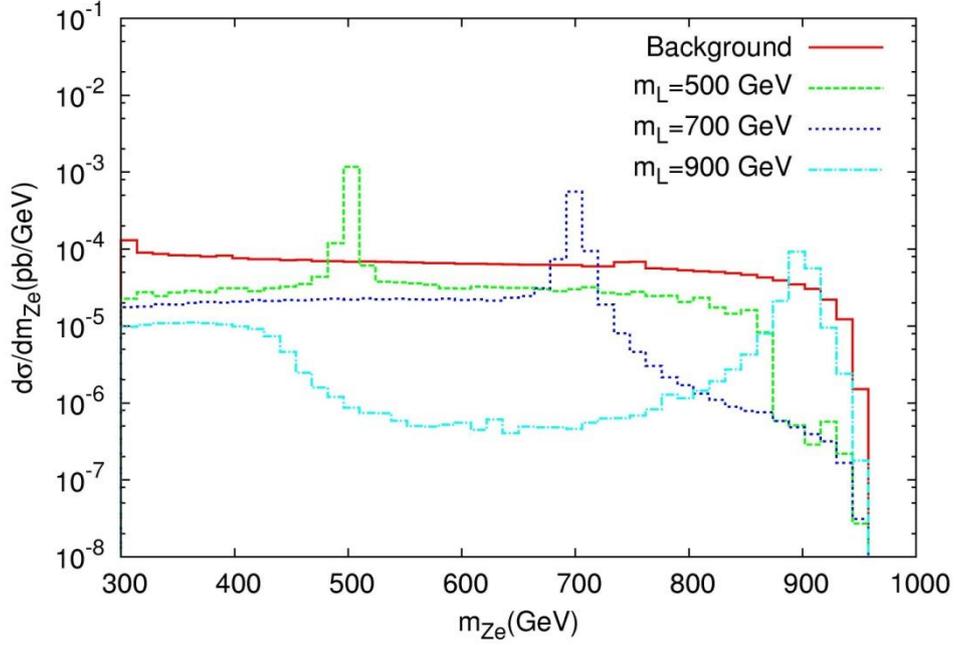

**FIG. 5:** The invariant mass distribution of $Ze$ in the final state at $\sqrt{s}$ = 1 TeV; signal with $m_L$ = 500 GeV, 700 GeV, 900 GeV and for the background ($e^-e^+ \to e^-Ze^+$) with $\kappa/\Lambda$ = 0.5 TeV$^{-1}$.

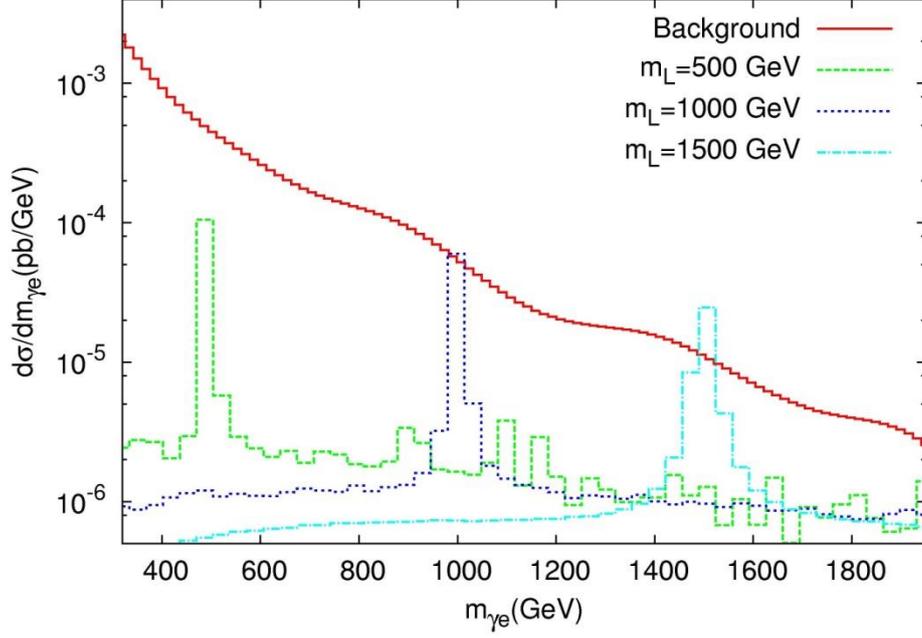

**FIG. 6:** The invariant mass distribution of $\gamma e$ in the final state at $\sqrt{s}$ = 3 TeV; signal with $m_L$ = 500 GeV, 1000 GeV, 1500 GeV and for the background ($e^-e^+ \to e^-\gamma e^+$) with $\kappa/\Lambda$ = 0.3 TeV$^{-1}$.

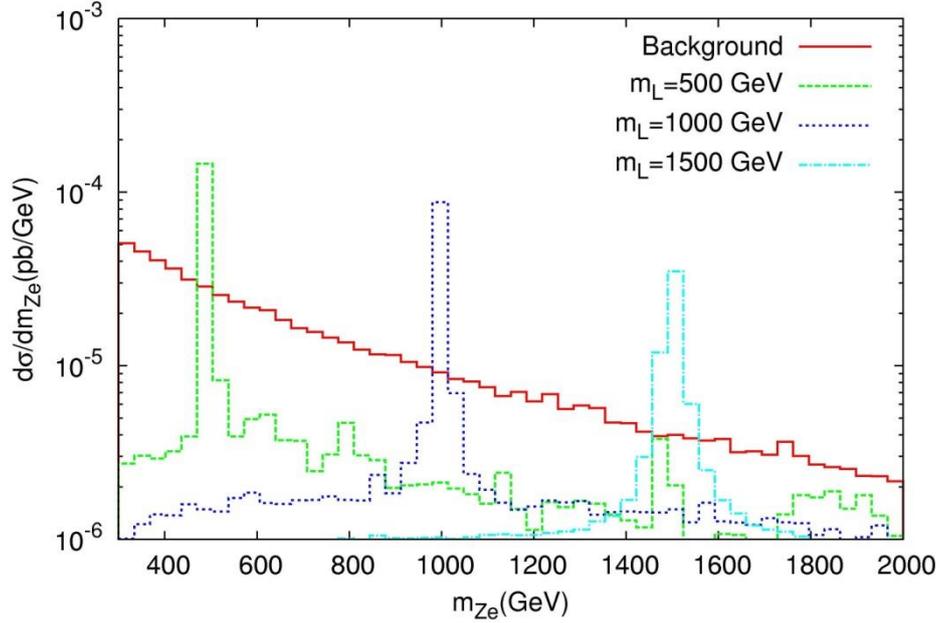

**FIG. 7:** The invariant mass distribution of $Ze$ in the final state at $\sqrt{s}$ = 3 TeV; signal with $m_L$ = 500 GeV, 1000 GeV, 1500 GeV and for the background ($e^-e^+ \to e^-Ze^+$) with $\kappa/\Lambda$ = 0.3 TeV$^{-1}$.

In order to obtain achievable values of the anomalous coupling strength, we require a statistical significance (SS) greater than 3 and at least 5 events per year of operation ($10^7$ s) as observation criteria. SS values are obtained from [24],

$$SS = \sqrt{2L^{int}\left[(\sigma_S + \sigma_B)\ln\left(1 + \frac{\sigma_S}{\sigma_B}\right) - \sigma_S\right]}$$

Here, $\sigma_S$ and $\sigma_B$ are signal and background cross sections, respectively. While calculating SS values, we consider the leptonic decays of $Z$ boson to muon pair with the branching value of 3.366% [25]. The obtained values of SS, $\sigma_S$ and $\sigma_B$, after the mentioned cuts, are presented as function of $m_L$ in Tables III and IV for the process $e^-e^+ \to e^-Ze^+$, in Tables V and VI for the process $e^-e^+ \to e^-\gamma e^+$ for the CLIC energy options.

**TABLE III:** The signal and SM background cross sections with ISR+Beamstruhlung for the process $e^-e^+ \to e^-Ze^+$ and SS depending on the heavy lepton masses with $\sqrt{s} = 1$ TeV, $L^{int} = 10^5$ pb$^{-1}$ and $\kappa/\Lambda = 0.5$ TeV$^{-1}$.

| $m_L$ (GeV) | $\sigma_S$ (pb) | $\sigma_B$ (pb) | SS |
|---|---|---|---|
| 300 | $2.26 \times 10^{-2}$ | $7.23 \times 10^{-4}$ | 19.8 |
| 500 | $1.63 \times 10^{-2}$ | $7.74 \times 10^{-4}$ | 15.7 |
| 700 | $8.08 \times 10^{-3}$ | $8.91 \times 10^{-4}$ | 9.2 |
| 900 | $2.12 \times 10^{-3}$ | $9.33 \times 10^{-4}$ | 3.2 |

**TABLE IV:** The signal and SM background cross sections with ISR+Beamstruhlung for the process $e^-e^+ \to e^-Ze^+$ and SS depending on the heavy lepton masses with $\sqrt{s} = 3$ TeV, $L^{int} = 10^5$ pb$^{-1}$ and $\kappa/\Lambda = 0.3$ TeV$^{-1}$.

| $m_L$ (GeV) | $\sigma_S$ (pb) | $\sigma_B$ (pb) | SS |
|---|---|---|---|
| 300 | $5.96 \times 10^{-3}$ | $3.98 \times 10^{-4}$ | 8.9 |
| 700 | $5.07 \times 10^{-3}$ | $3.02 \times 10^{-4}$ | 7.3 |
| 1100 | $2.17 \times 10^{-3}$ | $1.64 \times 10^{-4}$ | 5.2 |
| 1500 | $1.44 \times 10^{-3}$ | $1.73 \times 10^{-4}$ | 3.8 |
| 1900 | $1.04 \times 10^{-3}$ | $2.19 \times 10^{-4}$ | 2.8 |

**TABLE V:** The signal and SM background cross sections with ISR+Beamstruhlung for the process $e^-e^+ \to e^-\gamma e^+$ and SS depending on the heavy lepton masses with $\sqrt{s}$ = 1 TeV, $L^{int}$= $10^5$ pb$^{-1}$ and $\kappa/\Lambda$ = 0.5 TeV$^{-1}$.

| $m_L$ (GeV) | $\sigma_S$ (pb) | $\sigma_B$ (pb) | SS |
|---|---|---|---|
| 300 | $1.90 \times 10^{-2}$ | $2.05 \times 10^{-2}$ | 37.2 |
| 500 | $1.21 \times 10^{-2}$ | $3.71 \times 10^{-2}$ | 18.9 |
| 700 | $5.88 \times 10^{-3}$ | $2.43 \times 10^{-2}$ | 11.5 |
| 900 | $1.53 \times 10^{-3}$ | $1.27 \times 10^{-2}$ | 4.2 |

**TABLE VI:** The signal and SM background cross sections with ISR+Beamstruhlung for the process $e^-e^+ \to e^-\gamma e^+$ and SS depending on the heavy lepton masses with $\sqrt{s}$ = 3 TeV, $L^{int}$= $10^5$ pb$^{-1}$ and $\kappa/\Lambda$ = 0.3 TeV$^{-1}$.

| $m_L$ (GeV) | $\sigma_S$ (pb) | $\sigma_B$ (pb) | SS |
|---|---|---|---|
| 300 | $4.99 \times 10^{-3}$ | $5.11 \times 10^{-3}$ | 19.4 |
| 700 | $2.85 \times 10^{-3}$ | $2.78 \times 10^{-3}$ | 15.0 |
| 1100 | $1.56 \times 10^{-3}$ | $1.71 \times 10^{-3}$ | 10.6 |
| 1500 | $1.02 \times 10^{-3}$ | $3.40 \times 10^{-3}$ | 5.3 |
| 1900 | $7.44 \times 10^{-4}$ | $4.53 \times 10^{-3}$ | 3.4 |

Considering the observability criteria SS ≥ 3 (3$\sigma$), for the $e^-Ze^+$ final state, heavy leptons with masses up to 900 GeV for $\sqrt{s}$ = 1 TeV with $\kappa/\Lambda$ = 0.5 TeV$^{-1}$ and 1700 GeV for $\sqrt{s}$ = 3 TeV with $\kappa/\Lambda$ = 0.3 TeV$^{-1}$ can reachable. While, for $e^-\gamma e^+$ final state, these mass values changes as 900 GeV for $\sqrt{s}$ = 1 TeV with $\kappa/\Lambda$ = 0.5 TeV$^{-1}$ and 1900 GeV for $\sqrt{s}$ = 3 TeV with $\kappa/\Lambda$ = 0.3 TeV$^{-1}$. We plot Fig. 8 in order to obtain the achievable values of the anomalous coupling strength at low energy and luminosity options of the CLIC for 3 $\sigma$ observation limit.

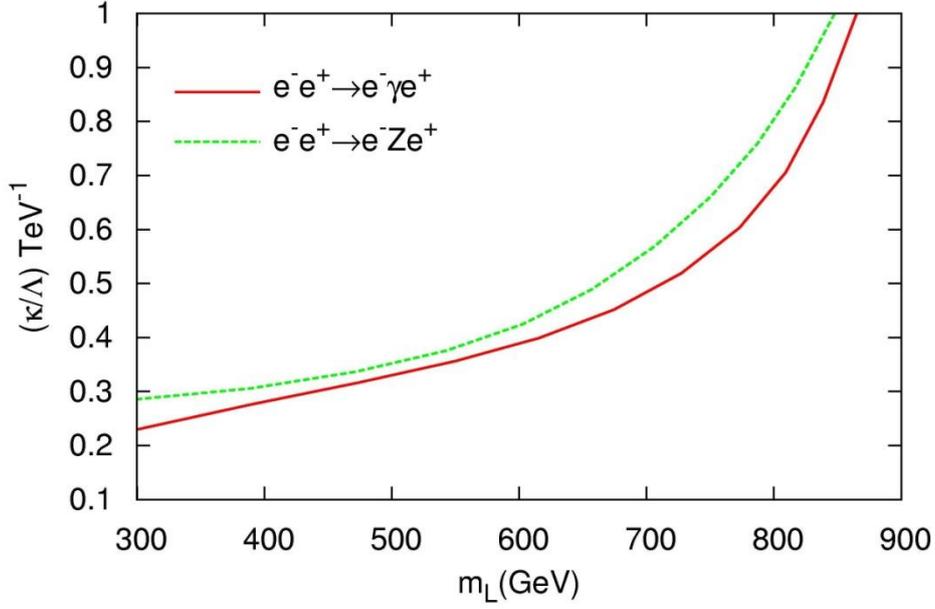

**FIG. 8:** The anomalous couplings as a function of $m_L$ for $3\sigma$ observation limit at $\sqrt{s}$ = 1 TeV with $L^{int}$ = 10 fb$^{-1}$.

One can see that, values as low as 0.3 TeV$^{-1}$ for $m_L$ = 450(400) GeV are reachable for $\kappa/\Lambda$ with $e^-\gamma(Z)e^+$ final state at $\sqrt{s}$ = 1 TeV with integrated luminosity of 10 fb$^{-1}$. In Fig. 9, the reachable values of $\kappa/\Lambda$ is plotted as function of $m_L$ for the standard luminosity ($L^{int}$ = 100 fb$^{-1}$) and energy ($\sqrt{s}$ = 3 TeV) of the CLIC.

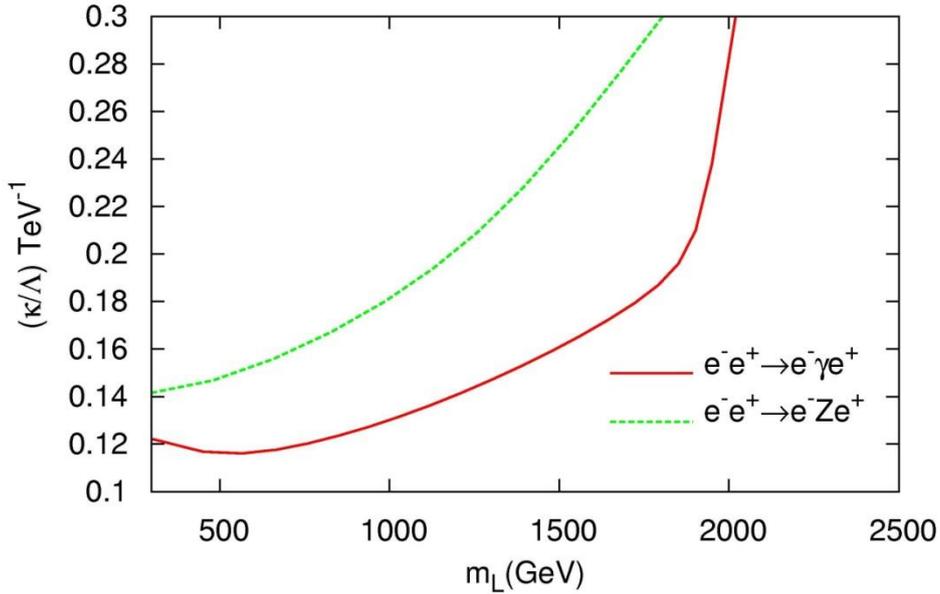

**FIG. 9:** The anomalous couplings as a function of $m_L$ for $3\sigma$ observation limit at $\sqrt{s}$ = 3 TeV with $L^{int}$ = 100 fb$^{-1}$.

As you can see from this figure, the reachable values of $\kappa/\Lambda$ are 0.12 (0.14) TeV$^{-1}$ for $m_L =$ 750(300) GeV with $e^-\gamma(Z)e^+$ final state at $\sqrt{s} = 3$ TeV and $L^{int} = 100$ fb$^{-1}$.

## IV. CONCLUSION

In this study, we investigate the effects of anomalous magnetic moment type interactions on single production and decays of charged heavy leptons at future linear colliders by taking into account ISR and beamstrahlung effects. We present the total cross section and experimental kinematical distributions of $e^-e^+ \to e^-\gamma e^+$ and $e^-e^+ \to e^-Ze^+$ processes for $\sqrt{s} = 1$ and 3 TeV options at the CLIC. Following the results from the numerical analysis, the sensitivity to the anomalous coupling $\kappa/\Lambda$ can reached down to 0.12 (0.3) TeV$^{-1}$ for $\sqrt{s} = 3(1)$ TeV at the CLIC. As a result, CLIC will provide a great facility to search for masses and anomalous magnetic dipole factors of charged heavy leptons.

**Acknowledgments**
We thank Orhan Çakır for his useful comments.